\documentclass[%
aip,
amsmath,amssymb,
reprint,%
]{revtex4-1}
\usepackage{physics}
\usepackage{graphicx}
\usepackage{dcolumn}
\usepackage{bm}
\usepackage{soul}

\usepackage[utf8]{inputenc}
\usepackage[T1]{fontenc}
\usepackage{mathptmx}
\usepackage{etoolbox}

\usepackage[toc,page]{appendix}

\makeatletter
\def\@email#1#2{%
	\endgroup
	\patchcmd{\titleblock@produce}
	{\frontmatter@RRAPformat}
	{\frontmatter@RRAPformat{\produce@RRAP{*#1\href{mailto:#2}{#2}}}\frontmatter@RRAPformat}
	{}{}
}%
\makeatother

\usepackage{color}
\usepackage{xcolor}
\definecolor{myblue}{RGB}{0,109,176}

\newcommand{\jrg}[1]{\textcolor{black}{#1}}

\usepackage[T1]{fontenc}
\usepackage{pgfplots}
\pgfplotsset{compat = newest}
\usetikzlibrary{arrows,intersections}
\usepackage{tikz-3dplot}
\usepackage{tikz}

\usepackage[mathscr]{euscript}
\usepackage{calligra}
\DeclareMathAlphabet{\mathcalligra}{T1}{calligra}{m}{n}
\DeclareFontShape{T1}{calligra}{m}{n}{<->s*[2.2]callig15}{}

\newcommand{\stability}{{\bm{A}}}

\newcommand{\ex}{{\bm{x}}}

\newcommand{\yu}{{\bm{u}}}
\newcommand{\brho}{{\bm{\varrho}}}

\newcommand{\logder}{{\boldsymbol{L}}}

\graphicspath{{../plots/}} 

\usepackage{tcolorbox}
\usepackage{multirow}
\usepackage{graphicx}
\definecolor{new_blue}{RGB}{9, 136, 232}
\newtcolorbox{mybox}[3][]
{%
	colframe = #2!25,
	colback  = #2!10,
	coltitle = #2!20!black,  
	title    = {#3},
	#1,
}
\tcbuselibrary{breakable}

\usepackage{pgfplots}
\pgfplotsset{compat = newest}
\usetikzlibrary{arrows,intersections}
\usepackage{tikz-3dplot}
\usepackage{tikz}

\makeatletter
\def\maketitle{
	\@author@finish
	\title@column\titleblock@produce
	\suppressfloats[t]}
\makeatother

\begin{document}
	
	\preprint{AIP/123-QED}
	
	\title[Classical Fisher information for differentiable dynamical systems]{Classical Fisher information for differentiable dynamical systems}
	\author{Mohamed~Sahbani}
	\affiliation{ 
		Department of Chemistry,\
		University of Massachusetts Boston,\
		Boston, Massachusetts 02125, USA} 
	\affiliation{Department of Physics,\
		University of Massachusetts Boston,\
		Boston, Massachusetts 02125, USA} 
	\author{Swetamber~Das}
	
	\affiliation{ 
		Department of Chemistry,\
		University of Massachusetts Boston,\
		Boston, Massachusetts 02125, USA} 
	\affiliation{Department of Physics,\
		University of Massachusetts Boston,\
		Boston, Massachusetts 02125, USA} 
	
	\author{Jason~R.~Green}
	\email{jason.green@umb.edu}
	
	\affiliation{ 
		Department of Chemistry,\
		University of Massachusetts Boston,\
		Boston, Massachusetts 02125, USA} 
	\affiliation{Department of Physics,\
		University of Massachusetts Boston,\
		Boston, Massachusetts 02125, USA} 
	
	\date{\today}

\begin{abstract}

Fisher information is a lower bound on the uncertainty in the statistical estimation of classical and quantum mechanical parameters.
While some deterministic dynamical systems are not subject to random fluctuations, they do still have a form of uncertainty:
Infinitesimal perturbations to the initial conditions can grow exponentially in time, a signature of deterministic chaos.
As a measure of this uncertainty, we introduce another classical information, specifically for the deterministic dynamics of \jrg{isolated, closed, or open} classical systems not subject to noise.
This classical measure of information is defined with Lyapunov vectors in tangent space, making it less akin to the classical Fisher information and more akin to the quantum Fisher information defined with wavevectors in Hilbert space.
Our analysis of the local state space structure and linear stability lead to upper and lower bounds on this information, giving it an interpretation as the net stretching action of the flow.
Numerical calculations of this information for illustrative mechanical examples show that it depends directly on the phase space curvature and speed of the flow. 

\end{abstract}

\maketitle

\section{Introduction}

Across science and engineering~\cite{watanabeClassicalEstimationTheory2014a,foxFiniteStateProjection2019}, Fisher information is widely used in the design of statistical experiments on both classical and quantum mechanical systems.
Loosely speaking, it is a measure of the amount of information that a data set contains about a statistical parameter.
In classical statistics, it was introduced by Fisher~\cite{fisherMathematicalFoundationsTheoretical1997} for experiment design and predicting the minimum achievable error in estimating physical quantities. It has since become an important metric in information geometry~\cite{Amari2016} and a part of physical theories~\cite{machtaParameterSpaceCompression2013}. 
For example, the classical Fisher information has been used to investigate phase transitions~\cite{Mikhail2011}, the emergence of criticality in living systems~\cite{Hidalgo2014} and collective synchronization~\cite{Kalloniatis2018}, and as a complexity measure in neural networks~\cite{Liang2020}. 
For quantum mechanical systems, the quantum Fisher information~\cite{helstromMinimumMeansquaredError1967,nielsenQuantumComputationQuantum2010} also guides measurements, characterizes the sensitivity of quantum states to system parameters, and sets limits on optimal estimation~\cite{parisQuantumEstimationQuantum2009}.
It is a key element in quantum speed limits~\cite{deffnerQuantumSpeedLimits2017, garcia-pintosUnifyingQuantumClassical2022} that has been applied to quantum criticality~\cite{yinQuantumFisherInformation2019}, quantum phase transitions~\cite{wangQuantumFisherInformation2014,marzolinoFisherInformationApproach2017}, coherence~\cite{luisFisherInformationGeneralized2012,yeQuantumCoherenceQuantum2020}, entanglement~\cite{obadaEntanglementEvaluationAtomic2010,liEntanglementDetectionQuantum2013}, and metrology~\cite{nielsenQuantumComputationQuantum2010}.

While classical and quantum theories are distinct, \jrg{their respective} Fisher information sets bounds on the estimation of parameters.
\jrg{Fisher information is} also an integral part of the geometric structure of the theories, defining a metric on the appropriate statistical manifold and, so, being related to measures of curvature~\cite{liuQuantumFisherInformation2019, sidhuGeometricPerspectiveQuantum2020}.
Consider a classical probability distribution $p(x;\theta)$ and the unbiased estimation of the parameter $\theta$.
The classical Fisher information $I^C_F$ is the expectation value $\langle (d_\theta\ln p)^2 \rangle$ of the square of the derivative of the log of the probability distribution with respect to $\theta$: by the Cram{\'e}r-Rao inequality, the variance $e^2$ in the unbiased estimation of $\theta$, $e^2I^C_F \geq 1$, cannot be smaller than $1/I^C_F$.
If the probability distribution is Gaussian with variance $\sigma^2$, the bound saturates and $I^C_F=\sigma^{-2}$.
\jrg{Sharply peaked Gaussian distributions will then be better estimates of the mean $\theta$ than broad distributions, and a given data set will have more information about the parameter.}
A more sharply peaked probability distribution indicates that the manifold has a higher local curvature \jrg{near the maximum}.
More precisely, $I^C_F \propto -\kappa_\text{max}/(1+(d_\theta\ln p_\text{max})^2)$, where $\kappa_\text{max}$ is the curvature of the distribution at its maximum.
From this intuition, Fisher information is \jrg{important to} the geometrical structure of statistical manifolds~\cite{nicholsonNonequilibriumUncertaintyPrinciple2018, Kim2021}.

Like its classical counterpart, the quantum Fisher information is the minimum possible error in the unbiased estimation of an unknown parameter. Here, though, the measurement is on a quantum state~\cite{yuQuantumFisherInformation2022} $\hat{\brho}$ that evolves unitarily under a Hamiltonian $\hat{H}$.
If $\hat{e}^2$ represents the variance of the quantum parameter being measured $\hat{\theta}$, then the quantum Fisher information sets the bound: $\hat{e}^2\mathcal{I}_F^Q\geq 1$.
For pure states, the quantum Fisher information $\mathcal{I}^Q_F=4\Delta \hat{H}^2$ is directly related to the variance of the Hamiltonian $\Delta \hat{H}^2$.
Among geometric formulations of quantum mechanics~\cite{Kibble1979, ashtekar1999, brody2001geometric, bengtsson_zyczkowski_2006,Anza2021,Anza2022a,Anza2022b}, the quantum Fisher information is a feature of density-matrix based information-geometric approaches~\cite{cafaroDecreaseFisherInformation2018}.
The quantum statistical manifold consisting of a collection of quantum states is endowed with a metric related to $\mathcal{I}_F^Q$, which defines the curvature of the manifold~\cite{sidhuGeometricPerspectiveQuantum2020}. 
On a manifold of pure density matrices, the quantum Fisher information sets the upper limit on the instantaneous rate of change of the absolute statistical distance~\cite{braunsteinStatisticalDistanceGeometry1994,kimInvestigatingInformationGeometry2018} between two pure quantum states.
This geometric perspective leads to quantum estimation and the comparison of quantum sensing protocols with $\mathcal{I}_F^Q$~\cite{loweLinkFisherInformation2022}.

In both classical and quantum theories, the Fisher information is also a part of some stochastic and quantum speed limits, upper bounds on how quickly observables evolve in time.
Both classical and quantum speed limits are rapidly developing fields with interesting parallels that include bounds set by the Fisher information~\cite{poggiDivergingQuantumSpeed2021, garcia-pintosUnifyingQuantumClassical2022, aghionThermodynamicSpeedLimits2023, pati2023exact}. 
Stochastic thermodynamic speed limits on observables set by the classical Fisher information, for example, apply to the fluxes of energy and entropy exchange~\cite{nicholsonTimeInformationUncertainty2020}.
While in isolated quantum systems, the Fisher information bounds the speed of the average energy, purity, and entropy~\cite{deffnerQuantumSpeedLimits2017}.
Recently, two of us~\cite{dasSpeedLimitsDeterministic2023} derived a classical speed limit on the growth rates of perturbations and rates of dissipation for classical deterministic systems.
One of these speed limits is set by a quantity that we can identify as a classical Fisher information.
\jrg{However,} this Fisher information $\mathcal{I}^C_F$ is a classical quantity, but it is distinct from the classical Fisher information based on probability distributions, $I_F^C$.
\jrg{
For non-Hamiltonian dynamics, it is defined in the tangent space of the state space. 
For many-particles systems in the position-momentum phase space, it is defined in the tangent space of a classical phase space point, making it a purely mechanical quantity.}

Here, we analyze this Fisher information for deterministic, differentiable dynamics to relate it to both local instability and state space curvature for classical, many-body systems.
For many-body Hamiltonian systems, there is evidence that the curvature of the potential energy landscape can affect the stability of dynamical trajectories and the fluctuations in the chaotic properties of fluids~\cite{dasSelfAveragingFluctuationsChaoticity2017,dasCriticalFluctuationsSlowing2019}.
Phase space curvature can also regularize the dynamics of barrier crossing, including the isomerization of atomic clusters~\cite{hindeChaoticDynamicsVibrational1993,greenSpacetimePropertiesGramSchmidt2009}, small molecules~\cite{greenCharacterizingMolecularMotion2011,greenChaoticDynamicsSteep2012}, and proteins~\cite{chekmarevAlternationPhasesRegular2019}.
The magnitude of (finite-time) Lyapunov exponents~\cite{pikovskyLyapunovExponentsTool2016} reflect this behavior and can even be directly related to the reactive flux over the barrier~\cite{revueltaTransitionStateTheory2017}.
These examples suggest the need for a better understanding of the relationship between phase space curvature and the speed of dynamical processes.

To address this need, we examine this Fisher information $\mathcal{I}^C_F$ for deterministic dynamical systems, showing that a classical density matrix theory~\cite{dasDensityMatrixFormulation2022} in Sec.~\ref{sec:classical-density-theory} leads naturally to its definition, Sec.~\ref{sec:TSFI}, and a relationship to the speed through state space.
Given the relationship of other forms of Fisher information to curvature, we corroborate the hypothesis that this new Fisher information is a measure of state space curvature with larger (smaller) values of $\mathcal{I}^C_F$ at a given state space point correspond to higher (lower) curvature, Sec.~\ref{sec:curvature}.
In addition, we derive upper and lower bounds on this Fisher information.
To gain physical insight, we compute these bounds for analytically tractable model oscillators, Sec.~\ref{sec:models}\jrg{, one conservative and one dissipative}.

\section{Classical density matrix theory} \label{sec:classical-density-theory}

We recently established a classical density matrix theory to describe the time evolution of infinitesimal perturbations in deterministic systems~\cite{dasDensityMatrixFormulation2022}.
This theory is based on the normalized time evolution of the perturbation vectors in the tangent space of classical systems and analogous to the density matrix formulation of quantum mechanics based on wavevectors in Hilbert space.
In this classical framework, the classical density matrix is mathematically similar to the metric tensor associated with the underlying phase space, which led to extensions of Liouville's theorem and equation for non-Hamiltonian systems~\cite{dasDensityMatrixFormulation2022}.
Local measures of dynamical instability and chaos also appear, but in classical (anti)commutators; for Hamiltonian dynamics, they are directly related to Poisson brackets.
We first summarize the parts of this theory that we will need.

Consider a differentiable dynamical system, $\dot{\ex} = \boldsymbol{F}(\ex)$, in which a point $\ex(t) := [x^1(t), x^2(t), \ldots , x^n(t)]^\top$ evolves in the $n$-dimensional state space. A small perturbation to that point, $\ket{\delta \ex}$, also evolves in time but according to the linearized dynamics:
\begin{align}\label{eq:linear-transformation} 
  \frac{d}{dt} \ket{\delta\ex(t)} = \stability[x(t)] \ket{\delta{\ex(t)}}.
\end{align}
The stability matrix, $\stability := \stability[x(t)] = \nabla{\boldsymbol{F}}$ has entries $\left(\stability\right)^{i}_{j}=\partial{\dot{x}^i(t)}/\partial{x^j(t)}$ and governs the time evolution of a perturbation vector $\ket{\delta \ex(t)}$. 
Here, we use Dirac's notation~\cite{diracPrinciplesQuantumMechanics1982} to represent a finite-dimensional column (row) vector with the ket (bra): $\ket{\delta{x^i(t)}} := [\delta{x^1(t)}, \delta{x^2}(t), \ldots , \delta{x^n(t)}]^\top$ in a real tangent space.

To define the classical density matrix and associated Fisher information, we use the norm-preserving dynamics of the unit perturbation vector $\ket{\delta{\yu(t)}}=\ket{\delta{\ex(t)}}/\lVert\ket{\delta \ex(t)}\rVert$.
Here, $\|.\|$ represents the $\ell^2$ norm: $\lVert\ket{\delta \ex(t)}\rVert := \langle \delta \ex|\delta \ex\rangle^{1/2}$.
The time evolution of this vector,
\begin{align}\label{eq:EOM-unit} 
  \frac{d}{dt}\ket{\delta \yu(t)} = \bar \stability\ket{\delta \yu(t)},
\end{align}
preserves the unit norm through the generator of the evolution $\bar\stability = \stability - \bra{\delta \yu(t)}\stability\ket{\delta \yu(t)} \mathbb{I}$. \jrg{The matrix $\mathbb{I}$ is the $n\times n$ identity matrix.}
The source/sink term $\bra{\delta \yu(t)}\stability\ket{\delta \yu(t)}$ is the instantaneous Lyapunov exponent\jrg{, which here counteracts any stretching or contraction and ensures the unit vector is normalized at all times.}
Overall, Eq.~\ref{eq:EOM-unit} \jrg{plays a role that} is similar to the Schr\"odinger equation from quantum mechanics with $\bar{\stability}$ playing the role of the Hamiltonian operator.
However, unlike the Hamiltonian, which has a Hermitian matrix representation, the real matrix $\bar{\stability}$ is generally not symmetric and depends on the unit perturbation vector at time $t$.

The classical density matrix associated with a unit perturbation vector $\ket{\delta \yu}$ is the dyadic product, $\brho(t) = \dyad{\delta \yu(t)}{\delta \yu(t)}$. 
Keeping in mind this symmetric matrix is composed of classical variables, we borrow the nomenclature of quantum mechanics and call this density matrix a \textit{pure} perturbation state.
It has the usual properties of a projection operator: it is positive semi-definite, $\brho \succeq 0$, with $\Tr\brho=1$ and $\brho = \brho^2$.
Partitioning the matrix $\stability$ into its symmetric $\stability_+ = (\stability + \stability^\top)/2$ and anti-symmetric $\stability_- = (\stability - \stability^\top)/2$ parts, the time evolution of the density matrix  $\brho$ is:
\begin{align}\label{eq:EOM-rho-1}
\frac{d}{dt}\brho = \{\stability_+, \brho\} + [\stability_-, \brho] - 2\langle\stability\rangle\brho.
\end{align}
The notation $\{\cdot\}$ and $[\cdot]$ indicate the anti-commutator and commutator, respectively, while $\langle \boldsymbol{X}\rangle :=\Tr(\boldsymbol{X}\brho)$ is the expectation value of $\boldsymbol{X}$ with respect to the classical density matrix $\brho$. 
Equation~\ref{eq:EOM-rho-1} is a classical analogue of the von-Neumann equation. Unlike the quantum version, this equation contains an anti-commutator along with a source/sink term $\langle \stability\rangle$ and is valid for differentiable deterministic systems, \jrg{not only those that are Hamiltonian}. 
From this equation of motion, we construct a classical Fisher information and consider its geometric interpretation in terms of phase space curvature and speed.

\section{Fisher information for classical pure states}\label{sec:TSFI}

To define this Fisher information on the first variations of the state variables (e.g., position and momentum), we start with Eq.~\ref{eq:EOM-rho-1} in the form~\cite{dasSpeedLimitsDeterministic2023}:
\begin{align}\label{eq:state-equation-of-motion}
\frac{d}{dt}\brho = \bar{\stability}\brho + \brho\bar{\stability}^\top.
\end{align}
Comparing to $d_t\brho :=\frac{1}{2}\left(\logder\brho + \brho \logder^\top\right)$, this equation of motion implicitly defines the logarithmic derivative $\logder$ of the density matrix, $\logder = 2\bar\stability$.
From the logarithmic derivative, we can then show the Fisher information $\mathcal{I}^C_F$ for a pure state $\brho$ is the variance of $\logder$ (App.~\ref{SM:Fisher}):
\begin{align}\label{eq:Fisher-info-by-ld}
  \mathcal{I}_F^{C} = \Delta{\logder^2}= 4\Delta{\stability^2}.
\end{align}
We define the variance for a matrix $\boldsymbol{X}$ as $\Delta \boldsymbol{X}^2 = \langle \boldsymbol{X}^\top \boldsymbol{X}\rangle - \langle \boldsymbol{X}^\top\rangle\langle\boldsymbol{X}\rangle$,
and use the fact that the expectation value of $\logder$ vanishes: $\langle \logder \rangle = 2\langle \bar{\stability}\rangle = 0$.
The same expression of $\mathcal{I}^C_F$ follows from the variance of the symmetric logarithmic derivative (App.~\ref{SM:Symmetric-logarithmic}).

The mathematical form of this Fisher information resembles the quantum Fisher information for pure quantum states.
For a pure quantum state evolving under unitary dynamics, the quantum Fisher information is $\mathcal{I}_F^Q = \Delta\hat{\logder}^2 = 4\Delta\hat{\boldsymbol{H}}^2/\hbar^2$ directly related to $\Delta\hat{\boldsymbol{H}}^2$, the variance of the Hamiltonian $\hat{\boldsymbol{H}}$ with respect to the pure state $\hat{\brho}$.
It derives from the symmetric logarithmic derivative $\hat{\logder}$ defined implicitly in $d_t\hat{\brho} = (\hat{\logder}\hat{\brho} + \hat{\brho} \hat{\logder})/2$ for a pure quantum state $\hat{\brho}$~\cite{sidhuGeometricPerspectiveQuantum2020}.
The expectation value $\langle\hat{\logder}\rangle$ also vanishes.

\subsection{Lower and upper bounds for pure states}

Because $\mathcal{I}^C_F$ for pure states is a variance, it is nonnegative, $\mathcal{I}_F^C\geq 0$, despite being computed with a perturbation vector in the tangent space of a point in state space.
\jrg{However, there is also an upper bound on this Fisher information:
At a given point in state space}, the spectrum of the stability matrix $\stability$ constrains the maximum value.
To derive this upper bound, we consider the definition of $\mathcal{I}^C_F$ in Eq.~\ref{eq:Fisher-info-by-ld} and write $\Delta \stability^2$ explicitly as:
\begin{align}\label{eq:Fisher-exp}
  \frac{1}{4}\mathcal{I}_F^C = \langle \stability^\top\stability\rangle - \langle\stability \rangle^2,
\end{align}
using the equality $\langle\stability\rangle = \langle \stability^\top\rangle$.
We use two secondary results to upper bound $\mathcal{I}_F^C/4$.

First, we recognize $\langle \stability^\top\stability\rangle$ as the Rayleigh quotient of $\ket{\delta \yu}$ with respect to the matrix $\stability^\top\stability$. 
The symmetric matrix $\stability^\top\stability$ is diagonalizable and $\brho$ is positive-semidefinite, so we can apply the min-max theorem~\cite{hornMatrixAnalysis1985} to get:
\begin{align}\label{eq:FI-bound00}
   \lambda_\text{min}(\stability^\top\stability)\leq  \frac{\Tr(\stability^\top\stability \brho)}{\Tr \brho} \leq\lambda_\text{max}(\stability^\top\stability).
\end{align}
The minimum, $\lambda_\text{min}(\stability^\top\stability)$, and maximum, $\lambda_\text{max}(\stability^\top\stability)$, eigenvalues of $\stability^\top\stability$ set bounds on $\langle \stability^\top\stability\rangle = \Tr(\stability^\top\stability \brho)$.
These inequalities have been rigorously proven~\cite{wangTraceBoundsSolution1986} for any two real matrices $\boldsymbol{X}$ and $\boldsymbol{Y}$ of order $n \times n$, provided $\boldsymbol{X}$ is symmetric and $\boldsymbol{Y}$ is positive semidefinite, $\lambda_\text{min}^{\boldsymbol{X}} \Tr(\boldsymbol{Y}) \leq \Tr(\boldsymbol{X}\boldsymbol{Y}) \leq \lambda_\text{max}^{\boldsymbol{X}} \Tr(\boldsymbol{Y})$.
Recognizing the dynamics of $\brho$ is trace-preserving, $\Tr\brho = 1$ for all times, the result here simplifies to (App.~\ref{SM:Fisher-Inequality}):
\begin{align}\label{eq:FI-bound0}
   \sigma^2_\text{min}(\stability)\leq  \langle \stability^\top\stability\rangle \leq\sigma^2_\text{max}(\stability),
\end{align}
with bounds set by the singular values $\sigma$ of $\stability$. 
The upper (lower) bound in Eq.~\ref{eq:FI-bound0} saturates when $\brho$ is composed of the eigenvector of $\stability^\top\stability$ corresponding to the maximum (minimum) eigenvalue.

Second, we recognize that the second term on the right hand side of Eq.~\ref{eq:Fisher-exp} is nonnegative, $\langle \stability\rangle^2 \geq 0$, 
and appears with a negative sign. 
Putting this fact together with the Rayleigh quotient in Eq.~\ref{eq:FI-bound00}, the bounds on $\mathcal{I}^C_F$ are:
\begin{align}\label{eq:FI-bound1}
  0\leq \frac{\mathcal{I}_F^C}{4} \leq\sigma^2_\text{max}(\stability).
\end{align}
This upper bound has a couple of useful features. Because the largest singular value of $\stability$ sets the maximum value, the bound is independent of the basis chosen for $\brho$. 
The upper bound also gives a mathematical interpretation of this Fisher information.
\jrg{Loosely speaking, s}ingular values are a measure of how much the matrix scales the space it acts on. 
So, we can interpret $\mathcal{I}^C_F$ as a measure of the extent to which we can estimate the stretching action of the linearized dynamics in the neighborhood of a point in state space. 

When does the Fisher information here saturate the upper and lower bounds?
If $\stability$ is a symmetric matrix, then $\langle\stability^2\rangle = \langle\stability\rangle^2$ in the direction of its eigenvectors.
Therefore, the Fisher information $\mathcal{I}_F^C$ vanishes.
In this case, the lower bound on $\mathcal{I}_F^C$ saturates.
For the upper bound to saturate, a perturbation vector $\ket{\delta \yu}$ needs to fulfill two conditions: (a) $\langle \stability \rangle = 0$, and (b) $\ket{\delta \yu}$ must be that singular vector of $\stability$ that corresponds to its maximum singular value.
When these conditions are met, we can see from Eqs.~\ref{eq:Fisher-exp} and ~\ref{eq:FI-bound0} that $\mathcal{I}_F^C = \sigma^2_\text{max}(\stability)$.
This happens, for example, in the simple harmonic oscillator, Sec.~\ref{subsec:SHO}.

A tighter, and nontrivial, lower bound is possible for basis sets where $\langle\stability\rangle$ is nonzero. 
To see how, we first recognize that $\langle \stability\rangle = \langle \stability_+\rangle = \Tr(\stability_+\brho)$ is an instantaneous Lyapunov exponent associated with a pure perturbation state $\brho$.
From Eq.~\ref{eq:FI-bound00}, this exponent is bounded by the extremal eigenvalues~\cite{DasGreen-spectral-bounds} of $\stability_+$. 
Then, the possible values of $\langle \stability_+\rangle^2$ belong to the interval $[0, \sigma_\text{max}^2(\stability_+)]$, where $\sigma_\text{max}(\stability_+)$ is the maximum singular value of the symmetric matrix $\stability_+$.
Now, using interval arithmetic, we obtain a potentially tighter lower bound on $\mathcal{I}^C_F$:
\begin{align}\label{eq:FI-bound2}
   \sigma^2_\text{min}(\stability)- \sigma^2_\text{max}(\stability_+)\leq \frac{\mathcal{I}_F^C}{4} \leq\sigma^2_\text{max}(\stability).
\end{align}
The bounds are independent of the basis chosen for $\brho = \dyad{\delta\yu}{\delta\yu}$. 
This lower bound will be tighter than zero for arbitrary differentiable dynamical systems with nonvanishing local Lyapunov exponents: specifically, when $\sigma^2_\text{min}(\stability) > \sigma^2_\text{max}(\stability_+)$.

\begin{figure}[t]
	\vspace{2mm}
	\begin{tikzpicture}
	\draw[draw=gray,-latex, ultra thick]
	(-1,4.05) to [out=20,in=135](5,5);
	\draw[draw=gray,-latex, ultra thick](0.46, 4.69) -- (1.0,4.95);
	
	\draw[draw=myblue,-latex,line width=0.5mm](2,5.4) -- (1.0,5.4+2.4);
	\node[align=left] at (0.1,6.95) {$\color{myblue}\ket{\boldsymbol{f}}=\logder\ket{\delta \yu(t)}$};
	
	\draw[draw=black,-latex,line width=0.5mm](2.0,5.38) -- (4,6.2); 
	\draw[fill=gray] (2.0,5.38) circle[radius=2pt];
	
	\draw[draw=black,-,line width=0.2mm](1.9,5.7) -- (2.2, 5.83);  
	\draw[draw=black,-,line width=0.2mm](2.2, 5.83) -- (2.338, 5.5); 
	
	
	\node[align=left] at (3.4,7.25) {$\mathcal{I}^C_F = \bra{\boldsymbol{f}}\ket{\boldsymbol{f}} = \|\logder\ket{\delta \yu(t)}\|^2$};
	\node[align=left] at (4.6,6.25) {$\ket{\delta \yu(t)}$};
	\node[align=left] at (4.3,5.0) {$\Gamma(t)$};
	\node[align=left] at (2.0,5.0) {$\ex(t)$};
	\end{tikzpicture}
	\caption{Geometric interpretation of the Fisher information $\mathcal{I}^C_F$: It is the $\ell^2$ norm of the vector $\ket{\boldsymbol{f}} = \logder\ket{\delta \yu}$, which is a result of applying a linear transformation $\logder$ to a perturbation vector $\ket{\delta \yu}$ at a given point $\ex(t)$ on a phase space trajectory $\Gamma(t)$. In the diagram, the vector $\ket{\delta \yu}$ aligns with the flow direction $\dot{\ex}$, and the angle between $\ket{\boldsymbol{f}}$ and $\ket{\delta \yu}$ is $\pi/2$ as $\langle \logder\rangle = 0$ at every point along a trajectory. The vector collinear with the direction of the flow is only one possible choice of $\ket{\delta \yu}$. \label{fig:vectors_on_orbit}} 
\end{figure}
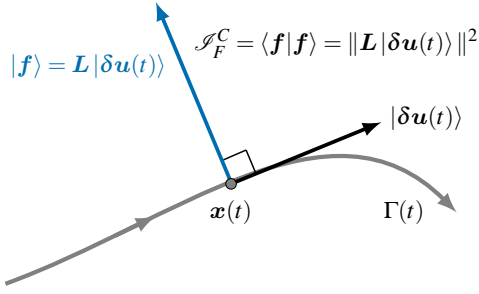

\subsection{Decomposition into phase space curvature and speed}\label{sec:curvature}

For a geometric interpretation of  $\mathcal{I}^C_F$, we use Eq.~\ref{eq:EOM-unit} to find (App.~\ref{SM:Fisher}):
\begin{align}\label{eq:FI-mag}
	 \mathcal{I}^C_F = 4\|\ket{\delta \dot{\yu}}\|^2 = \|\logder \ket{\delta \yu}\|^2=\bra{\delta \yu}\logder^\top\logder\ket{\delta \yu}. 
\end{align}
We consider an infinitesimal perturbation vector $\ket{\delta \yu}$ in the tangent space at a point $\ex(t)$ on the state space trajectory $\Gamma(t)$.
The logarithmic derivative $\logder$ is a linear transformation on $\ket{\delta {\yu}}$ producing another vector $\ket{\boldsymbol{f}} = \logder\ket{\delta {\yu}}$.
The vector $\ket{\boldsymbol{f}}$ is the \textit{Fisher vector} and the matrix $\logder^\top\logder$ is the Fisher matrix.
Its $\ell^2$ norm gives the tangent space Fisher information at $\ex(t)$ for the pure state of perturbation $\brho = \dyad{\delta \yu(t)}{\delta \yu(t)}$.
Because the projection of $\ket{\delta {\yu}}$ on $\ket{\boldsymbol{f}}$ given by $\langle \logder\rangle = \bra{\delta {\yu}}\logder\ket{\delta {\yu}}$ vanishes, the angle between an arbitrary tangent vector and Fisher vector $\ket{\boldsymbol{f}}$ is $\pi/2$ at all points along a trajectory. For a system with Hamiltonian $H$, the Fisher vector aligns with $\nabla H$.

A natural choice for a perturbation vector is in the flow direction $\dot\ex$ at a state-space point, Fig.~\ref{fig:vectors_on_orbit}.
For an infinitesimal time $\delta t$, the perturbation in the flow direction is $\ket{\delta \ex} = \ket{\dot \ex} \delta t$. 
Once normalized, $\ket{\delta \yu} = \bra{\delta \ex}\ket{\delta \ex}^{-1/2} \ket{\delta \ex}$, we write it explicitly for the flow in the $n$-dimensional phase space:
\begin{align}\label{eq:perturbed-pure-state}
     \ket{\delta \yu_{\dot \ex}}=\frac{[{\dot{x}_1(t)}, {\dot{x}_2}(t), \ldots , {\dot{x}_n(t)}]^\top}{\sqrt{{\dot{x}^2_1(t)} + {\dot{x}^2_2(t)} + \ldots + {\dot{x}^2_n(t)}}} =\begin{pmatrix}
	u_1\\\vdots\\
	u_n\end{pmatrix},
\end{align}
where $u_i = \dot{x}_i/ \sqrt{\sum_{i=1}^n \dot x^2_i}$ so that the normalization condition $\sum_{i=1}^nu^2_i = 1$ is satisfied at all times.
The subscript $\dot{\ex}$ indicates the flow direction.
We will use this vector to relate the Fisher information here to the state space curvature.

A classical trajectory experiences the local curvature as it evolves through state space.
Geometrically, the curvature at a point on the trajectory is the rate of change of the unit perturbation vector $\ket{\delta \yu_{\dot \ex}}$ along the tangential direction to the flow $\dot{\ex}$.
The curvature of a trajectory segment of length $s$ is (App.~\ref{SM:curvature}):
\begin{align}
\kappa(t) = \norm{\frac{d\ket{\delta \yu_{\dot\ex}}}{ds}},
\end{align}
where $s(t) =  \int_{t_0}^{t} \norm{\ket{\dot \ex}} dt'$ for a trajectory given by integrating the flow $\dot \ex$.
Applying the chain rule, the curvature becomes:
\begin{align}
  \kappa(t) = \dot s^{-1}\norm{\ket{\delta \dot \yu_{\dot\ex}}} = \norm{\ket{\dot \ex}}^{-1}\norm{ \ket{\delta  \dot\yu_{\dot \ex}}},
\end{align}
with $\dot s = \norm{\ket{\dot \ex}}$ by the fundamental theorem of calculus. From Eq.~\ref{eq:FI-mag}, the Fisher information in the flow direction is
\begin{align}\label{eq:F-curvature}
 \mathcal{I}^C_F(\dot{\ex}) = 4\kappa^2\norm{\ket{\dot \ex}}^2
\end{align}
\jrg{is determined by the local curvature and the phase speed along a trajectory.}
\jrg{This form of the Fisher information is specific to the direction of the flow and independent of whether the dynamics are conservative or dissipative.}

The upper and lower bounds in the previous section immediately apply to this form of the Fisher information.
To evaluate the bounds and to elucidate the connection between $\mathcal{I}^C_F(\dot{\ex})$ with phase space curvature, we consider paradigmatic model mechanical systems.

\section{Model systems}\label{sec:models}

\subsection{Free particle}

The classical free particle has a nonnegative kinetic energy and no potential energy.
It is linear system with Hamiltonian $H(q, p)= p^2/2m$ and equations of motions $\dot {q} = p/m$ and $\dot{p} = 0$.
The particle thus moves in a straight line with a constant momentum $p$.
A perturbation in the direction of the flow is time independent: 
\begin{align*}
      \ket{\delta{\yu}_{\dot \ex}}=\frac{[{\dot{q}(t)}, {\dot{p}(t)}]]^\top}{\sqrt{{\dot{q}^2(t)}+ {\dot{p}^2(t)}}} = \pm\begin{pmatrix}
	1\\
	0\end{pmatrix}, \qquad 
\end{align*}
where $\pm$ indicates the direction of the phase space velocity vector $[{\dot{q}(t)}, {\dot{p}(t)}]]^\top$.
The time derivative of $\ket{\delta{\yu}_{\dot\ex}}$ vanishes.
As a consequence, the Fisher information, $\mathcal{I}^C_T= 0$, from Eq.~\ref{eq:FI-mag} also vanishes, reflecting the curvature-less geometry of the free particle phase space.

\subsection{Linear harmonic oscillator}\label{subsec:SHO}

As an example with a controllable phase space structure, we consider the one-dimensional linear harmonic oscillator~\cite{goldstein2002classical} with mass $m$, frequency $\omega$, and Hamiltonian $H(q,p) =p^2/2m + m\omega^2q^2/2$.
Its equations of motion are:
\begin{align}
	\dot {q} = p/m, \quad \text{and} \quad \dot{p} = -m\omega^2q.
\end{align}
For a given total energy, it is well known that the harmonic oscillator has an elliptical phase space.
The ellipses are a convenient feature for illustrating the dependence of the Fisher information on phase space curvature.
In fact, in suitable dimensionless coordinates $Q$ and $P$, we can introduce a single parameter to control the orientation and major axes of the ellipses, and, so, the variations in curvature an oscillator will sample along a dynamical trajectory.
We define the generalized coordinate $Q$ and the conjugate momentum $P$ through:
\begin{align}\label{eq:unit-less-coordinates}
  q = {q}_mQ,\qquad \text{and} \qquad   p = {p}_m P.
\end{align}
The coordinates $Q$ and $P$ are made dimensionless by the constant factors $q_m$ and $p_m$ with units of length $[L]$ and momentum $[MLT^{-1}]$.
This coordinate transformation requires that the Poisson brackets be $\{Q(q,p),P(q,p)\} = 1$ and $\{Q(q,p),Q(q,p)\} = \{P(q,p),P(q,p) = 0\}$.
Together, these lead to the condition ${q_m}{p_m} = 1$. We will use this relation to define a de-dimensionalization parameter.

To write the transformed Hamiltonian, we introduce a parameter $a:= q_m/p_m$, with dimensions $[TM^{-1}]$.
This parameter controls the orientation of the phase space ellipses through the relative contributions of the kinetic and potential energies to the Hamiltonian.
Larger (smaller) values of $a$ will amplify the potential (kinetic) energy term and the curvature of phase space trajectories around $q=0$ ($p=0$).
Using $a$, the Hamiltonian is:
\begin{align}\label{eq:Hamiltonian-a}
\mathcal{H}(Q,P,a) ={q_m}{p_m}\omega\left(\frac{P^2}{2ma\omega} + \frac{ma\omega{Q}^2}{2}\right).
\end{align}
The quantity ${q_m}p_m{\omega}$ has the dimensions of energy $[ML^2T^{-2}]$ and the terms inside the parenthesis in Eq.~\ref{eq:Hamiltonian-a} are dimensionless. 
The non-dimensionalized equations of motion become
\begin{align}\label{eq:unit-less-SHO}
\dot{Q}= \frac{P}{ma}, \qquad     \text{and} \qquad     \dot{P}=-ma{\omega^2}Q
\end{align}
with the solution:
\begin{align}\label{eq:HO-sol-PQ}
Q(t) = B\cos(\omega{t} + \phi) , \quad
P(t) = -ma\omega B\sin(\omega{t} + \phi).
\end{align}
The amplitude $B$ is set by the initial condition. The stability matrix for the system in Eq.~\ref{eq:unit-less-SHO} is:
\begin{figure}[t]
	\includegraphics[width=0.8\columnwidth]{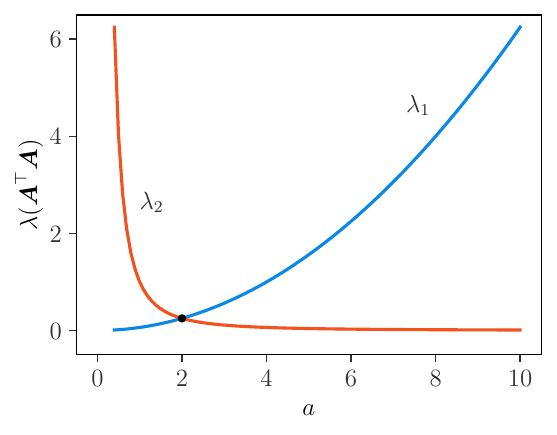}
	\caption{Two eigenvalues of the matrix $\stability^\top\stability$, $\lambda_\text{1}= (ma\omega^2)^2$ and $\lambda_\text{2} =(ma)^{-2}$ varies with the parameter $a$. We set $m = 1$ and $\omega = 0.5$ here. These eigenvalues gives extremal bounds on $\mathcal{I}^{C}_F$ for the simple harmonic oscillator. There is a spectral singularity at $a = (m\omega)^{-1} = 2$ (black dot) where both eigenvalues are 1/4.  At this point,  $\mathcal{I}^{C}_F$ attains a basis and time independent value $4\omega^2$ which corresponds to phase space trajectories of constant curvature. The minimum and maximum bounds on $\mathcal{I}^{C}_F$ for $a$ < 2 are $\lambda_2$ and $\lambda_1$, respectively. The bounds switch for $a > 2$ because $\lambda_1$ is now greater than $\lambda_2$. \label{fig:lambda_with_a}}
\end{figure}
\begin{align}\label{eq:SHO-stabilty matrix}
    \stability = \omega\begin{pmatrix}
	0&(ma\omega)^{-1}\\
	-ma\omega&0\end{pmatrix}.
\end{align}
By manipulating the dimensions in this way, the stability matrix and the Fisher information have dimensions of inverse time squared.
The Fisher information depends, in part, on the positive semi-definite symmetric matrix: 
\begin{align}\label{eq:symmetric-matrix}
    \stability^\top\stability = {\omega}^2\begin{pmatrix}
	(ma\omega)^2&0\\
	0&(ma\omega)^{-2}\end{pmatrix},
\end{align}
with the dimensions of ${\omega}^2$ and eigenvalues:
\begin{align}
	\lambda_\text{1}= (ma\omega^2)^2 \quad \text{and} \quad \lambda_\text{2} =(ma)^{-2}.
\end{align}
These eigenvalues are the square of singular values $\sigma^2$ of $\stability$. 
The parameter $a$ scales the mass $m$ and ensures the stability matrix has dimensions of frequency squared.
The potential energy $V = {q_m{p_m}}ma\omega^2Q^2/2$ is a function of $a$ and has the dimensions of energy.

\begin{figure*}[t]
\includegraphics[width=0.7\textwidth]{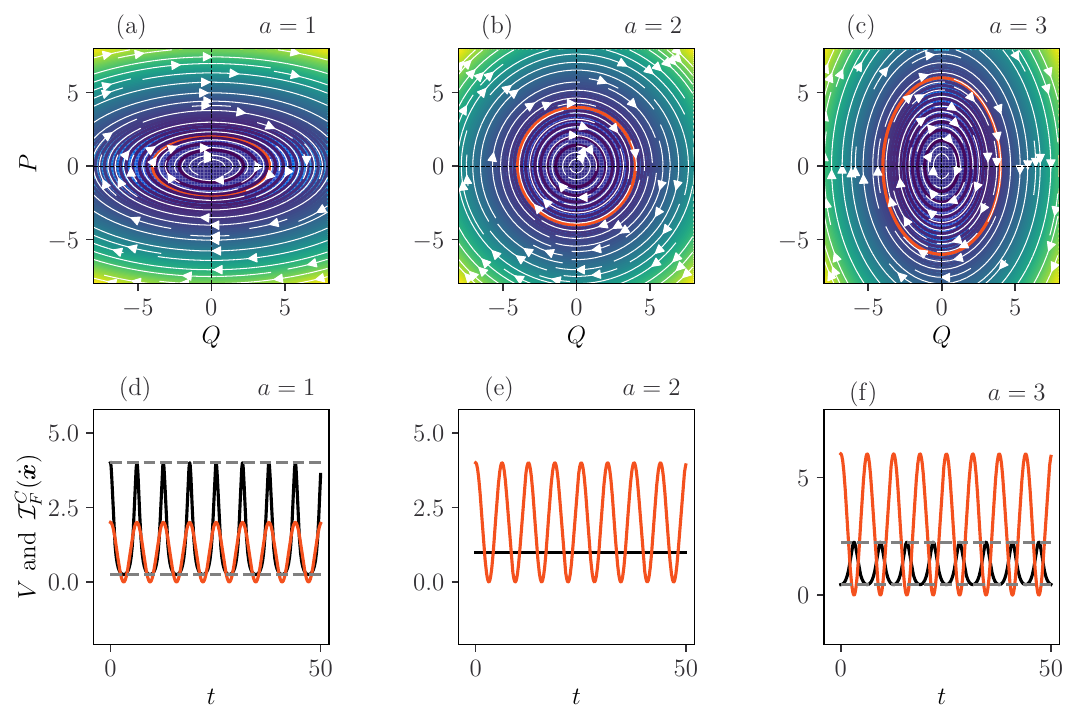}
\caption{\textit{Simple harmonic oscillator. ---}  Phase space of the simple harmonic oscillator for (a) $a = 1$, (b) $a=2$, and (c) $a = 3$. In each case, one trajectory (shown in red) is chosen to compute potential energy $V$ and Fisher information $\mathcal{I}^C_F(\dot{\ex})$. Time evolution of $V$ (red) and $\mathcal{I}^{C}_F(\dot{\ex})$ (in black) for (d) $a = 1$, (e) $a = 2$, and (f) $a = 3$. The bounds on $\mathcal{I}^{C}_F(\dot{\ex})$ are marked by horizontal dashed line in gray. Here, mass $m = 1$ and frequency $\omega =0.5$. \label{fig:SHO-phase-curve-potential-Fisher}
}
\end{figure*}

Now, for an arbitrary perturbation vector $\ket{\delta{\yu}} = [u, v]^\top$, the Fisher information for the simple harmonic oscillator (App.~\ref{SM:Fisher-undamped-damped}) is:
\begin{align}\label{eq:explicit-FI} 
    \mathcal{I}^C_F=4\Delta \stability^2 = 4\omega^2\left(ma\omega{u}^2 + \frac{v^2}{ma\omega}\right)^2.
\end{align}
According to Eq.~\ref{eq:FI-bound0} (App.~\ref{SM:Fisher-bounds}), it is bounded by the extremal eigenvalues $\lambda_\text{min}$ and $\lambda_\text{max}$, of $\stability^\top\stability$:
\begin{align}\label{eq:bounds2}
  4\lambda_\text{min}\leq \mathcal{I}^C_F\leq 4\lambda_\text{max}.
\end{align}
These bounds on the Fisher information are independent of the choice of perturbation vector.
For $a < (m\omega)^{-1}$, the minimum and maximum bounds are $\lambda_\text{min}$ = $\lambda_2$ and $\lambda_\text{max}$ = $\lambda_1$.
The matrix $\stability^\top\stability$ has an exceptional point, a spectral singularity at the point $a = (m\omega)^{-1}$ where the two eigenvalues are degenerate and the corresponding eigenvectors coalesce.
In Fig.~\ref{fig:lambda_with_a}, this point is at $a = 2$.
At this point, $\mathcal{I}^C_F$ has a constant value determined by $4\omega^2$ which is one for $\omega = 0.5$.
The bounds switch for $a > (m\omega)^{-1}$, $\lambda_\text{min}$ = $\lambda_1$ and $\lambda_\text{max}$ = $\lambda_2$.
Figure~\ref{fig:lambda_with_a} shows $\lambda_1$ and $\lambda_2$ as a function of $a$ for $m = 1$ and $\omega = 0.5$.
For the initial condition at time $t = 0$, we chose a maximum amplitude of the oscillator by $Q(t_0) = 4$ and $P(t_0) = 0$.
These initial conditions, together with $m$ and $\omega$, also set the total energy of the system.

\subsection*{Upper and lower bounds on $\mathcal{I}^C_F(\dot{\ex})$ and their saturation}

As we have seen, Fisher information in the direction of the flow is directly related to the curvature of the trajectory. This information is readily computed for the simple harmonic oscillator, including the lower and upper bounds. \jrg{The upper bound saturates at the turning points on the trajectory where the potential energy is at its highest point. The lower bound saturates at the minimum potential energy, when kinetic energy is at its highest.}
In the flow direction $\dot \ex$, we start with the expression for $\mathcal{I}^C_F$, which takes the form (App.~\ref{SM:Fisher-alternatives}):
\begin{align} 
  \mathcal{I}^C_F (\dot{\ex})  =4\omega^2\left(\frac{ma\omega}{1 + [(ma\omega)^2 - 1]\frac{V}{E}}\right)^2,
\end{align}
where $0\leq V/E \leq 1$. From this expression, the Fisher information depends on the potential $V$ for fixed energy $E$, $a$, $m$, and $\omega$. Since $ma\omega$ and $V/E$ are dimensionless, it has dimensions of $\omega^2$.

Bounds on $\mathcal{I}^C_F(\dot{\ex})$ follow from:
\begin{align}\label{eq:FII}
    \mathcal{I}^C_F(\dot{\ex}) =4\omega^2 \left(\frac{ma\omega}{1 + \left(\left(ma\omega\right)^2 -1\right)\cos^2(\omega{t} + \phi)}\right)^2,
\end{align}
which we derive in App.~\ref{SM:Fisher-bounds}.
The extremal values of the cosine function lead to two cases:
\begin{align}\label{eq:bounds1}
\frac{4}{m^2a^2}&\leq{ \mathcal{I}^C_F(\dot{\ex})}\leq4m^2a^2{\omega^4}, & & \text{for} \quad a\leq\frac{1}{m\omega} \\
4m^2a^2{\omega^4}&\leq{ \mathcal{I}^C_F(\dot{\ex})}\leq\frac{4}{m^2a^2}, & & \text{for} \quad a\geq\frac{1}{m\omega}.
\end{align}
For $a<1/m\omega$, the minimum and maximum bounds saturate at phase angles $\omega t + \phi = (l+1)\pi/2$ and $\omega t + \phi = l\pi$, respectively, for $l \in \{0, 1, 2, \cdots\}$.
At these phase angles, for $a>1/m\omega$, it is the maximum and minimum bounds that are saturated, respectively.
The bounds saturate because at these angles $\langle \stability\rangle = 0$ and thus $\langle \stability\rangle$ does not contribute to $\mathcal{I}^C_F(\dot{\ex})$.

To illustrate these bounds, we choose three representative values of $a \in \{1, 2, 3\}$ that determine the orientation of the phase space ellipse.
We choose $m = 1$ and $\omega = 0.5$ such that the exceptional point is located at $a = 2$.
Panels (a), (b), and (c) in Fig.~\ref{fig:SHO-phase-curve-potential-Fisher} show the phase space trajectories for each $a$.
The geometry of the phase space is an oblate ellipse at $a = 1$, circle at $a = 2$, and prolate ellipse at $a=3$.
As shown in panels (d), (e), and (f) in Fig.~\ref{fig:SHO-phase-curve-potential-Fisher}, the Fisher information $\mathcal{I}^C_{F}(\dot\ex)$ and the potential energy $V = q_m{p_m}ma\omega^2Q^2/2$ oscillate as the phase point traces out a trajectory.

For $a = 1$, the oscillations of $\mathcal{I}^C_F(\dot{\ex})$ are in phase with those in  $V$, Fig.~\ref{fig:SHO-phase-curve-potential-Fisher}(d).
The periodic variation in $\mathcal{I}^C_F(\dot{\ex})$ saturates the upper and lower bounds at its maximum and minimum values, respectively.
In addition, we can see that $\mathcal{I}^C_F(\dot{\ex})$ correlates with the local curvature of the trajectory.
The points of largest local curvature occur at the turning points in Fig.~\ref{fig:SHO-phase-curve-potential-Fisher}(a) where the potential energy $V$ is maximum. Figure~\ref{fig:SHO-phase-curve-potential-Fisher}(d) shows that at these points, $\mathcal{I}^C_F(\dot\ex)$ is the largest, saturating the upper bound. Similarly, at points of lowest curvature which corresponds to $V = 0$, the local curvature is the smallest, saturating the lower bound.

As shown in Figure~\ref{fig:SHO-phase-curve-potential-Fisher}(e), at the exceptional point $a=2$, the phase space trajectory traces out a circle of radius $R = \sqrt{2mE}$ and constant curvature $R^{-1}$.
In this case, $\mathcal{I}^C_F(\dot{\ex})$ is constant and both bounds saturate.
The potential energy does not share this signature of the constant curvature.
The case $a = 3$ shown in Fig.~\ref{fig:SHO-phase-curve-potential-Fisher}(f) is complementary to $a=1$.
The potential energy is out of phase with $\mathcal{I}^C_F(\dot{\ex})$.
The phase space structure is instead a prolate ellipse as can be seen in panel (c), with a maximum local curvature at $Q= 0$ and a minimum at the turning points.
Once again, $\mathcal{I}^C_F(\dot{\ex})$ correlates with the local curvature at all points along the trajectory and its maximum (minimum) saturates the upper (lower) bound.

\subsection*{Relation between $\mathcal{I}^C_F(\dot{\ex})$ and the energy and period}

In the case of the simple harmonic oscillator, the Fisher information can also be expressed as a function of the Hamiltonian. Combining the Hamiltonian in  Eq.~\ref{eq:F-curvature}, Eq.~\ref{eq:Hamiltonian-a}, and Eq.~\ref{eq:explicit-FI}, we get (App.~\ref{SM:F-curvature}):
\begin{align}\label{eq:Fisher-curvature}
  \mathcal{I}^C_F(\dot{\ex})= \left(16\omega^2\frac{\mathcal{H}}{{q_m}{p_m}}\right)^\frac{2}{3}\kappa^{\frac{4}{3}} ,
\end{align}
where $\mathcal{H}/q_mp_m$ has the dimensions of frequency $\omega$ and $\kappa$ is the local curvature. 
From this equation, we conclude that for a constant energy given by $\mathcal{H}$, the Fisher information $\mathcal{I}^C_F(\dot{\ex}) \propto \kappa^\frac{4}{3}$. 

One interpretation of the square root of the Fisher information (parameterized by time) is as a speed of the evolution across the statistical manifold defined by $\brho$~\cite{deffnerQuantumSpeedLimits2017,sidhuGeometricPerspectiveQuantum2020}.
Because the Fisher information has dimensions of inverse time squared, we can also relate it to the natural frequency of the oscillator.
Leaning on our interpretation, we derive an expression for the period of the harmonic oscillator (App.~\ref{SM:Period}):
\begin{align}
  T = 2 \int_{s_0}^{s}\frac{\kappa (s')}{ \sqrt{\mathcal{I}^{C}_F(\dot{\ex})}} ds'
\end{align}
When $a = 2$, the phase space is a circle, Fig.~\ref{fig:SHO-phase-curve-potential-Fisher}(b), with radius $\kappa = R^{-1}$.
The Fisher information is constant $\mathcal{I}^C_F(\dot{\ex}) = 4\omega^2$.
The arc length of the circle $s$ can then be expressed in terms of the angle such that $s = R\theta$.
We then get
\begin{align}
  T = 2 \int_{0}^{2\pi}\frac{1}{R}\frac{1}{ \sqrt{4\omega^2}}Rd{\theta} =\frac{2\pi}{\omega},
\end{align}
recovering the well known period of the simple harmonic motion.

\begin{figure*}[t]
\includegraphics[width=0.7\textwidth]{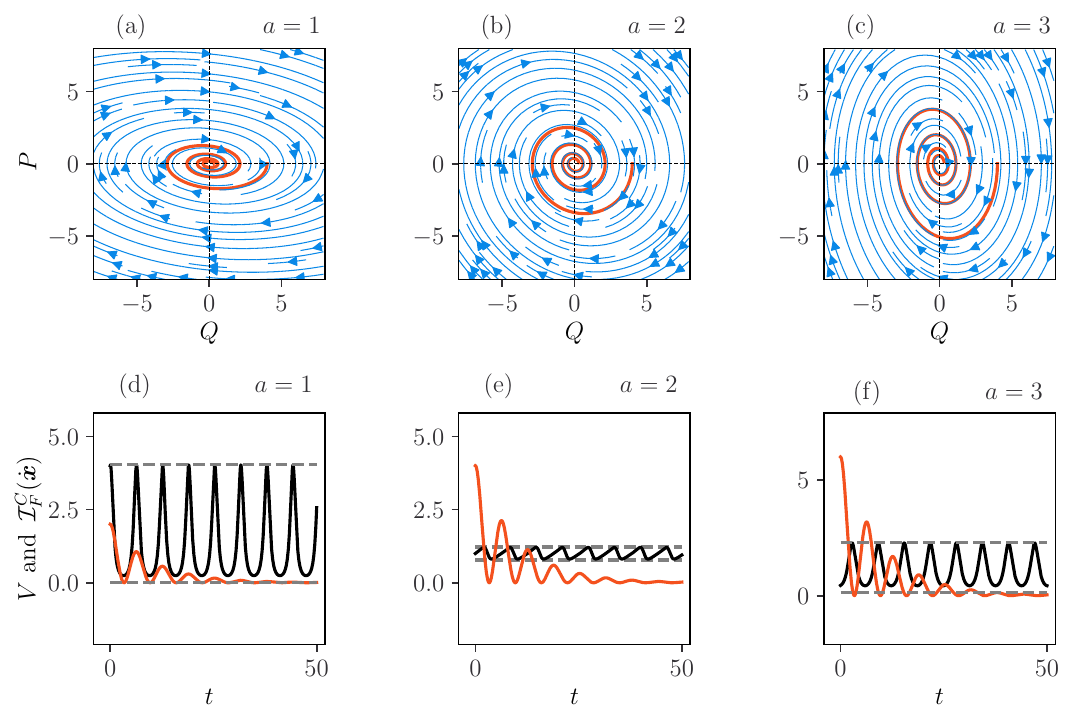}
\caption{\textit{Underdamped oscillator.---} The phase space of the underdamped harmonic oscillator for (a) $a = 1$, (b) $a=2$, and (c) $a = 3$. In each case, one trajectory (shown in color red in (a), (b), and (c)) is chosen to compute potential energy $V$ and Fisher information $\mathcal{I}^C_F(\dot{\ex})$. The three bottom panels show the time evolution of $V$ (red) $\mathcal{I}^C_F(\dot{\ex})$ (black) for (d) $a = 1$, (e) $a = 2$, and (f) $a = 3$ for the chosen trajectories. The bounds on $\mathcal{I}^C_F(\dot{\ex})$ are marked by horizontal dashed line (gray). For computations, we have set mass $m = 1$, frequency $\omega =0.5$ and $\gamma = 0.1$.\label{fig:DHO-under-damped-Fisher}}
\end{figure*}

\subsection{Damped harmonic oscillator}

\jrg{For comparison to the conservative dynamics of the simple harmonic oscillator, we analyzed the dissipative dynamics of the damped harmonic oscillator.
In dissipation dynamics, the phase space contracts over time, which we expect to reduce the curvature and speed along a trajectory and, so, the Fisher information.}

For the damped harmonic oscillator, the equations of motion are: 
\begin{align}\label{eq:Damped-equations-of-motion}
\dot{q}= \frac{p}{m}\qquad          \dot{p}=-m{\omega^2}q - \frac{\gamma}{m}p,
\end{align}
where $\gamma$ is the positive friction coefficient~\cite{goldstein2002classical}. Again using the parameter $a$ to non-dimensionalize, we get:
\begin{align}\label{eq:unit-less-DHO}
\dot{Q}= \frac{P}{ma}, \qquad     \text{and} \qquad     \dot{P}=-ma{\omega^2}Q - \frac{\gamma}{m}P.
\end{align}
In these coordinates, the stability matrix is now,
\begin{align}\label{eq:damped-stability-matrix}
    \stability = \omega\begin{pmatrix} 
	0&(ma\omega)^{-1}\\
	-ma\omega&-\gamma(m\omega)^{-1}\end{pmatrix}.
\end{align}
As for the harmonic oscillator, we can use the symmetric matrix,
\begin{align}\label{eq:damped-symmetric-matrix}
    \stability^\top\stability = {\omega}^2\begin{pmatrix}
	(ma\omega)^2&\gamma{a}\\
	\gamma{a}&(1 + \gamma^2 a^2)(ma\omega)^{-2}\end{pmatrix},
\end{align}
to construct the Fisher information.

The analytical form of Fisher information $\mathcal{I}^C_F = 4\Delta \stability^2$ for a general perturbation vector  $\ket{\delta{\yu}}=[u, v]^\top$ for the damped harmonic oscillator is (App.~\ref{SM:Fisher-under-critically-damped}):
\begin{align}\label{eq:damped-FI}
\mathcal{I}^C_F=4\omega^2\left(ma\omega{u}^2 + \frac{\gamma{a}}{ma\omega}uv + \frac{v^2}{ma\omega}\right)^2.
\end{align}
For the perturbation vector in the flow direction $\dot{\ex}$, it becomes (App.~\ref{SM:Fisher-under-critically-damped}):
\begin{align}\label{eq:Final-damped-FI}
  \mathcal{I}^C_F(\dot{\ex})=4\frac{\omega^6}{\norm{\dot{\boldsymbol{X}}}^4}\left(ma\omega{Q}^2 + \frac{\gamma{a}}{ma\omega}QP + \frac{P^2}{ma\omega}\right)^2,
\end{align}
where $\norm{\dot{\boldsymbol{X}}} = \sqrt{\dot P^2 + \dot Q^2}$ uses the equations of motion in Eq.~\ref{eq:unit-less-DHO}.
For $\gamma = 0$, we recover the expressions of Fisher information for the harmonic oscillator. 

As is well known, depending on the damping constant $\gamma$, frequency $\omega$, and mass $m$, there are three distinct cases -- underdamped, critically damped, and overdamped. 
We will consider these cases next, setting $m = 1$ and $\omega = 0.5$.

\begin{figure*}[t]
	\includegraphics[width=0.7\textwidth]{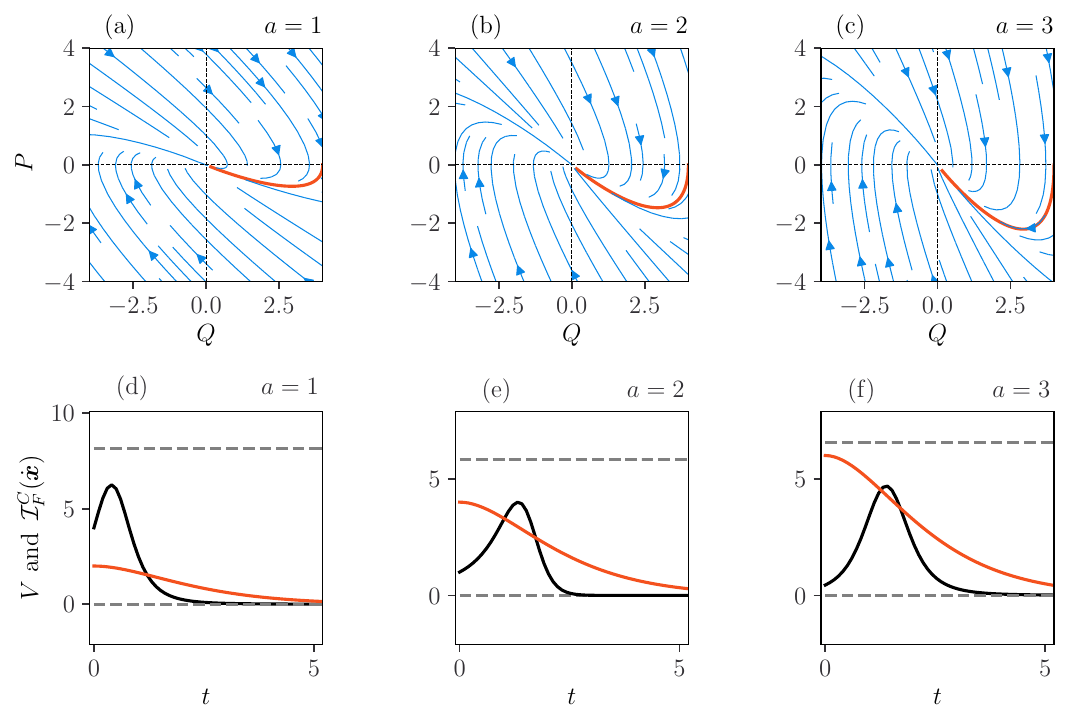}
	\caption{\textit{Critically damped oscillator. ---} The top panels show the phase space of the critically-damped harmonic oscillator for (a) $a = 1$, (b) $a=2$, and (c) $a = 3$. In each case, one trajectory (red in (a), (b), and (c)) is chosen to compute potential energy $V$ and Fisher information $\mathcal{I}^C_F(\dot{\ex})$. The three bottom panels show the time evolution of $V$ (red) $\mathcal{I}^C_F(\dot{\ex})$ (black) for (d) $a = 1$, (e) $a = 2$, and (f) $a = 3$. The bounds on $\mathcal{I}^C_F(\dot{\ex})$ are marked by horizontal dashed line (gray). For computations, we have set mass $m = 1$, frequency $\omega =0.5$, and $\gamma = 1.0$.\label{fig:DHO-critically-damped-Fisher}}
\end{figure*}

\medskip
\noindent\textit{Underdamped oscillator.--} \jrg{The angular frequency for the damped harmonic motion is $\omega_d = \sqrt{\omega^2 - (\gamma/2m)^2}$.} 
For small damping $\gamma$, with $\gamma/2m < \omega$, the system oscillates with frequency $\omega_d < \omega$ with this motion dampened exponentially over time.
The solutions to the equations of motion in Eq.~\ref{eq:unit-less-DHO} are
\begin{align*}
Q(t) &=  C\exp(-\frac{\gamma}{2m}t)\cos(\omega_d{t} + \phi), \\
P(t) &= - Ce^{-\frac{\gamma}{2m}t}\left(\frac{\gamma{a}}{2}\cos(\omega_d{t} + \phi) + ma\omega_d\sin(\omega_d{t} + \phi)\right),
\end{align*}
where $\phi = -\arctan[(\frac{\gamma{a}}{2} + \frac{P(t_0)}{Q(t_0)})/(ma\omega_d)]$, $C$ is a constant, $Q(t_0)$ and $P(t_0)$ are the initial conditions.
The Fisher information for underdamped oscillator is (App.~\ref{SM:Fisher-under-critically-damped}):
\begin{align}
 \mathcal{I}^C_F(\dot{\ex}) =4\omega^2 \left(\frac{ma{\omega}{{{\omega^{2}_d}}}{C^2}e^{-\frac{\gamma}{m}t}}{\norm{\dot{\boldsymbol{X}}}^2}\right)^2.
\end{align}
Fig.~\ref{fig:DHO-under-damped-Fisher} shows the phase space trajectories, Fisher information, and potential energy for the three representative values of $a$.
Unlike the harmonic oscillator, the point $a = 2$ is no longer an exceptional point. As a consequence of damping, phase trajectories spiral inwards to the phase space origin; see panels (a), (b), and (c) in Fig.~\ref{fig:DHO-under-damped-Fisher}. The local curvature of these trajectories varies in all three values of $a$ as indicated by changing shapes of the spirals and their orientation about the $(Q, P)$ axes.

As in the simple harmonic oscillator, the Fisher information $\mathcal{I}^C_F(\dot{\ex})$ exhibits oscillatory behavior in response to changes in local curvature along the spiraling trajectories.
To illustrate this behavior, we chose a representative trajectory for each of the three values of $a$.
These trajectories are red in Fig.~\ref{fig:DHO-under-damped-Fisher} (a)-(c).
For these trajectories, the potential energy $V$ and the Fisher information $\mathcal{I}^C_F(\dot{\ex})$ are shown for (d) $a= 1$, (e) $a= 2$, and (f) $a= 3$.
However, $\mathcal{I}^C_F(\dot{\ex})$ does not decay over time.
In this case, the effect damping is neutralized because the vector $\ket{\delta \yu_{\dot\ex}}$ is \textit{normalized} at all times (as long as the dynamics persist in the system).
The potential energy $V$ decays exponentially due to damping, as expected.
As a consequence, the largest (smallest) local curvature of the trajectory does not exactly correspond to local maxima (minima) in $V$.
The oscillations in $\mathcal{I}^C_F(\dot{\ex})$ and $V$ are no longer in phase because the spiral geometry of the trajectories is not symmetric about the $(Q,P)$ axes.

The numerical calculations confirm the bounds on $\mathcal{I}^C_{F}(\dot{\ex})$ in Fig.~\ref{fig:DHO-under-damped-Fisher}(d)-(f).
The upper bound is still given by the largest eigenvalue of $\stability^\top\stability$ which does not saturate as $\langle\stability\rangle$ never vanishes.
To compute the lower bound, we use the singular value of $\stability$ and $\stability_+$ according to Eq.~\ref{eq:FI-bound2}.
In panel (d), the lower bound computed from Eq.~\ref{eq:FI-bound2} is not useful as, in this case, $\sigma^2_\text{min}(\stability)$ does not exceed $\sigma^2_\text{max}(\stability_+)$.
However, the lower bound is tighter than zero in panels (e) and (f) because $\sigma^2_\text{min}(\stability) > \sigma^2_\text{max}(\stability_+)$ at all phase space points.

\medskip
\noindent\textit{Overdamped and critically damped oscillator.---}
Our results for the overdamped and critically damped dynamics are qualitatively similar, so we present our numerical results for the latter.
The oscillator exhibits critical damping when $\gamma/(2m) =\omega$, i.e., $\omega_d = 0$.
In this case, the equations of motion have the solution:
\begin{align*}
Q(t) &= \left(B + Ct\right)e^{-\frac{\gamma}{2m}t}, \\
P(t) &=  \left(maC -\frac{\gamma{a}}{2}B + \frac{\gamma{a}}{2}Ct\right)e^{-\frac{\gamma}{2m}t},
\end{align*}
where $B$ is a dimensionless constant and $C$ is a constant with dimensions of frequency $[T^{-1}]$.
Both are set by initial conditions $Q(t_0)$ and $P(t_0)$. Their ratio is:
\begin{align}
\frac{B}{C} = \frac{ma}{\frac{\gamma{a}}{2} + \frac{P_0}{Q_0}},
\end{align}
where $B = Q_0$, $C = \omega{Q_0} + \frac{P_0}{ma}$, and $\omega = \gamma/2m$.
In this case, the Fisher information takes the form (App.~\ref{SM:Fisher-under-critically-damped}):
\begin{align}
  \mathcal{I}^C_F(\dot{\ex}) =4\omega^2 \left(\frac{\sqrt{ma{\omega}}{\boldsymbol{C}}e^{-\frac{\gamma}{2m}t}}{\norm{\dot{\boldsymbol{X}}}}\right)^4,
\end{align}
where $\norm{\dot{\boldsymbol{X}}} = \sqrt{\dot Q^2 + \dot P^2}$.

When overdamped or critically damped, the oscillator does not show oscillations.
Its phase trajectories dampen exponentially until motion ceases.
The oscillator's phase space for $a \in \{1, 2, 3\}$ are shown in panels (a), (b), and (c), respectively, of Fig.~\ref{fig:DHO-critically-damped-Fisher}.
To compute trajectories, we set $m = 1$, $\gamma = 1$, and $\omega = 0.5$.
As before, we chose one representative trajectory for each of the three values of $a$ and show the corresponding potential energy and Fisher information in the panels (d), (e) and (f) of Fig.~\ref{fig:DHO-critically-damped-Fisher} for $a = 1$, 2, and 3, respectively.
 
In each of these cases, the phase space trajectory is a curve with one local maximum before reaching the fixed point.
The corresponding Fisher information has a single peak and vanishes when the trajectory reaches the origin.
Intuitively, there is no Fisher information when the system is static and at a fixed point.
A higher peak in $\mathcal{I}^C_F(\dot{\ex})$ indicates a higher curvature in the corresponding trajectory.
The potential energy decays, as expected.
The bounds on $\mathcal{I}^C_F(\dot{\ex})$ are indicated by horizontal dashed line in gray.
The upper bounds are the largest eigenvalue of $\stability^\top\stability$ for the given $a$.
The lower bound is zero as  $\sigma^2_\text{min}(\stability)$ never exceeds $\sigma^2_\text{max}(\stability_+)$.

\subsection{Fisher information $\mathcal{I}^C_F$ in higher dimensional systems}

\jrg{In the preceding subsection, we obtained closed form expressions for $\mathcal{I}^C_F$ and demonstrated its relation with phase space curvature for low-dimensional conservative and dissipative systems.
For higher dimensional systems that are not amenable to analytical treatment, $\mathcal{I}^C_F$ is numerically computable.
For instance, the Fisher information is computable from molecular dynamics simulations.
The stability matrix has been computed for atomic clusters~\cite{hindeChaoticDynamicsVibrational1993,greenSpacetimePropertiesGramSchmidt2009}, small molecules~\cite{greenCharacterizingMolecularMotion2011,greenChaoticDynamicsSteep2012}, simple liquids~\cite{PhysRevA.38.473,dasSelfAveragingFluctuationsChaoticity2017}, and proteins~\cite{chekmarevAlternationPhasesRegular2019}.
The results here are also applicable to classic dynamical systems, such as the Lorenz-Fetter model, for which one can compute the numerical trajectory, the stability matrix $\stability$, and the unit perturbation vector in the flow direction, $\ket{\delta \yu_{\dot \ex}}$ (from Eq.~\ref{eq:perturbed-pure-state}).
With these ingredients, $\ket{\delta \yu_{\dot \ex}}$ and $\stability$, one can compute the Fisher information in the flow direction $\dot{\ex}$ using Eq.~\ref{eq:Fisher-info-by-ld}: $\mathcal{I}_F^{C}(\dot{\ex}) = 4\Delta{\stability^2} = 4 \bra{\delta \yu_{\dot \ex}}\stability^\top\stability\ket{\delta \yu_{\dot \ex}} - 4 \bra{\delta \yu_{\dot \ex}}\stability\ket{\delta \yu_{\dot \ex}}^2$.
This algorithm can also be used to calculate $\mathcal{I}^C_F$ for arbitrary perturbation vector $\ket{\delta \yu}$ at any phase space point~\cite{dasSpeedLimitsDeterministic2023}.}

\section{Conclusions} \label{sec:conclusions}

We introduced a Fisher information for the differentiable dynamics of classical systems.
While classical, it is distinct from the well-known classical Fisher information associated with probability distributions that is used in the estimation of statistical parameters.
\jrg{Its mathematical form is more akin to the quantum Fisher information, and, for many-particle systems, it is a purely mechanical quantity that depends on positions and momenta.
Analyzing this Fisher information $\mathcal{I}^C_F$ led to several interpretations.
As a part of a recent classical density matrix theory~\cite{dasDensityMatrixFormulation2022}, this information measure is directly related to the fluctuations in local dynamical stability for pure states.
For these pure states, it is also a measure of the speed at which the system progresses through the space of classical density matrices. 
The square of the singular value of the stability matrix sets an upper bound and gives an interpretation as the net stretching action of the flow. 
Considering the} direction of the flow, the Fisher information depends on the curvature of phase space; A higher (lower) value of $\mathcal{I}^C_F$ indicates greater (smaller) curvature.
To illustrate this interpretation, we analyzed the conservative and damped harmonic oscillator to obtain analytical expressions of $\mathcal{I}^C_F$ and its bounds.
Overall, this classical information measure is computable for deterministic, differentiable systems, is a mechanical counterpart to the classical Fisher information in statistics, and offers a new perspective on how the time-evolution of classical systems depends on the geometry of state space.

\begin{acknowledgments}
This material is based upon work supported by the National Science Foundation under Grant No. 2124510. 
\end{acknowledgments}

\section*{Appendix}

\appendix
\renewcommand{\theequation}{\thesubsection\arabic{equation}}
\setcounter{equation}{0}
\renewcommand{\thesubsection}{\Alph{subsection}}
\subsection{Derivation of Fisher information} \label{SM:Fisher}

The equation of motion of a pure state defined by the perturbation vector $\ket{\delta u}$ is:
\begin{align}
\frac{d}{dt}\ket{\delta \yu} = \bar\stability\ket{\delta \yu}.
\end{align}
A pure state of perturbation defined as $\brho=\dyad{\delta \yu}{\delta \yu}$ evolves in time as
\begin{align}\label{SM:EOM-rho}
\frac{d}{dt}\brho = \bar\stability\brho + \brho\stability^\top.
\end{align}
Using the logarithmic derivative $\logder:=2\bar\stability$, the Fisher information is:
\begin{align}
\mathcal{I}^C_F &:= \Delta \logder^2 =\langle \logder^\top\logder\rangle \nonumber\\
&= 4\langle\bar\stability^\top\bar\stability\rangle = 4\left(\langle\stability^\top\stability\rangle - \langle\stability \rangle^2\right) = 4\Delta\stability^2.
\end{align}
Here, the average $\langle\cdot\rangle$ is with respect to $\brho$: $\langle \boldsymbol{X}\rangle = \Tr(\boldsymbol{X}\brho)$.
Alternatively, we can use $d_t\ket{\delta \yu}$ to express $ \mathcal{I}^M_F$ as:
\begin{align}
\mathcal{I}^C_F &= 4\langle\bar\stability^\top\bar\stability\rangle = 4 (\bra{\delta \yu} \bar\stability^\top)\left(\bar\stability\ket{\delta \yu}\right) \nonumber\\
& = 4\norm{\frac{d}{dt}\ket{\delta \yu}}^2.
\end{align}

\subsection{Symmetric logarithmic derivative and Fisher information}\label{SM:Symmetric-logarithmic}
\setcounter{equation}{0}
The equation of motion of a pure state is:
\begin{align}
\frac{d\brho}{dt} = \frac{d\brho^2}{dt} &= \frac{d\brho}{dt}\brho +  \brho \frac{d\brho}{dt} = \frac{\left(2d_t\brho\right)\brho +  \brho\left(2d_t\brho\right)}{2}.
\end{align}
Because $\brho$ is symmetric, its time derivative $d_t\brho$ is also symmetric.
We then obtain the symmetric logarithmic derivative $\logder_S = 2d_t\brho$.
The Fisher information is the expectation value of $\logder_S$ for $\brho$:
\begin{align}
\mathcal{I}^C_F &= \langle{\logder_S}^2\rangle = \langle\left(2d_t\brho\right)^2\rangle= 4\langle\left(d_t\brho\right)^2\rangle.
\end{align}
From Eq.~\ref{SM:EOM-rho},
\begin{align}
\left(\frac{d\brho}{dt}\right)^2 
&= \left(\bar{\stability}\brho + \brho\bar{\stability}^\top\right)\left(\bar{\stability}\brho + \brho\bar{\stability}^\top\right)\\\nonumber
&= \bar{\stability}\brho\bar{\stability}\brho + \bar{\stability}\brho\brho\bar{\stability}^\top + \brho\bar{\stability}^\top\bar{\stability}\brho + \brho\bar{\stability}^\top\brho\bar{\stability}^\top.
\end{align}
Using $\brho^2 = \brho$ and $\langle \bar\stability\rangle = 0$, the expectation value with respect to $\brho$,
\begin{align}
\left\langle\left(\frac{d\brho}{dt}\right)^2\right\rangle = \langle \bar\stability^\top \bar\stability\rangle = \Delta\stability^2,
\end{align}
which gives the Fisher information for a pure state:
\begin{align}
\mathcal {I}^{C}_F = 4\Delta \stability^2.
\end{align}

\subsection{Fisher information inequality of singular values}\label{SM:Fisher-Inequality}
\setcounter{equation}{0}
We partition the stability matrix $\stability$ into its symmetric and anti-symmetric parts, $\stability = \stability_++\stability_-$, and notice that:
\begin{align}
\langle \stability\rangle = \langle\stability_+\rangle,
\end{align}
since $\langle \stability_-\rangle = \Tr(\stability_-\brho) = 0$, i.e., the trace of the product of a symmetric matrix and an anti-symmetric matrix vanishes.
As explained in the main text, according to the min-max theorem and norm preservation $\Tr \brho = 1$, the minimum and maximum eigenvalues of $\stability_+$ bound $\langle\stability_+\rangle$ :
\begin{align} 
\lambda_\text{min}\left(\stability_+\right)\leq \langle{\stability_+}\rangle\leq\lambda_\text{max}\left(\stability_+\right),
\end{align}
where $\lambda_\text{min}$ and $\lambda_\text{max}$ are minimum and maximum eigenvalues of $\stability_+$ respectively. The square of $\langle\stability_+\rangle$ is bounded from below by zero and above by the square of its maximum singular values:
\begin{align}
0\leq \langle\stability_+\rangle^2\leq \sigma^2_\text{max}(\stability_+).
\end{align}
Similarly, $\langle \stability^\top \stability\rangle$ is bounded by,
\begin{align} 
\lambda_\text{min}\left(\stability^\top \stability\right)\leq \langle{\stability^\top \stability}\rangle\leq\lambda_\text{max}\left(\stability^\top \stability\right),
\end{align}
the minimum and maximum eigenvalues of $\stability^\top \stability$: $\lambda_\text{min}\left(\stability^\top \stability\right)$ and $\lambda_\text{max}\left(\stability^\top \stability\right)$, respectively.
These eigenvalues are the square of minimum and maximum singular values $\sigma_\text{min}$ and $\sigma_\text{max}$ of $\stability$.

These bounds extend to the Fisher information $\mathcal{I}^C_F$.  
Using the bounds on the terms on the right hand side of $\tfrac{1}{4}\mathcal{I}^C_F = \langle \stability^\top\stability\rangle - \langle\stability_+\rangle^2$, we obtain:
\begin{align} 
\sigma^2_\text{min}(\stability)-\sigma^2_\text{max}\left(\stability_+\right)\leq\tfrac{1}{4}\mathcal{I}^C_F\leq\sigma^2_\text{max}(\stability).
\end{align}

\subsection{The connection between curvature and Fisher information} \label{SM:curvature}
\setcounter{equation}{0}
The curvature $\kappa$ of a state space trajectory of length $s$ is related to the perturbation vector $\ket{\delta \yu_{\dot{\ex}}}$ in the flow direction $\dot\ex$:
\begin{align}
\kappa(t) = \norm{\frac{d}{d s}\ket{\delta \yu_{\dot{\ex}}}}.
\end{align}
From the chain rule, we get
\begin{align}
\kappa(t) &= \norm{\frac{d}{dt}\ket{\delta{\yu}_{\dot{\ex}}}}\norm{\frac{ds}{dt}}^{-1}
\end{align}
and recognize $ds/dt= \norm{d{\ket{\ex}}/dt} = \norm{\ket{\dot{{\ex}}}}$.
Using this result in Eq.~\ref{eq:FI-mag}, the Fisher information becomes:
\begin{align}
  \mathcal{I}^{C}_F(\dot{\ex}) = 4\norm{\frac{d}{dt}\ket{\delta{{\yu}_{\dot{\ex}}}}}^2 = 4\kappa^2\norm{\ket{\dot{{\ex}}}}^2.
\end{align}

\subsection{Fisher information for the damped harmonic oscillator}\label{SM:Fisher-undamped-damped}
\setcounter{equation}{0}
The stability matrix of the harmonic oscillator,
\begin{align}\stability = \omega
\begin{pmatrix}
0&1/ma\omega\\
-ma\omega &-\gamma/m\omega
\end{pmatrix},
\end{align}
has an expectation value of $\stability$ with respect to a pure state $ \ket{\delta{\yu}} = (u, v)^\top$:
\begin{align}
\langle\stability\rangle &= \bra{\delta u}\stability \ket{\delta u}
= \frac{v}{ma}\left(u -\left(\left(ma\omega\right)^2u + \gamma{a}v\right)\right).
\end{align}
Using $u^2 + v^2 = 1 $ and some algebra, we get
\begin{align}
\langle\stability\rangle^2   &=  \frac{v^2}{\left(ma\right)^2}\left(u -\left(\left(ma\omega\right)^2u + \gamma{a}v\right)\right)^2.
\end{align}
Similarly, we obtain:
\begin{align}\nonumber
\langle\stability^\top\stability\rangle  & = \omega^2\left((ma\omega)^2u^2 + 2\gamma{a}uv + \frac{1 + \left(\gamma{a}\right)^2 }{(ma\omega)^2}v^2\right)\\
&=  \frac{1}{\left(ma\right)^2}\left(\left((ma\omega)^2u + \gamma{a}v\right)^2 + v^2\right).
\end{align}
Now, we can calculate the variance of the stability matrix $\stability$ from the equation $\Delta \stability^2 = \langle \stability\stability^\top\rangle - \langle \stability^\top\rangle\langle\stability\rangle$:
\begin{align}
\begin{split}
\Delta \stability^2 &=  \omega^2\left(ma\omega{u^2} + \frac{\gamma{a}}{ma\omega}uv  + \frac{v^2}{ma\omega} \right)^2.
\end{split} 
\end{align}
Finally, the Fisher information for the damped harmonic oscillator is:
\begin{align}
\mathcal{I}^C_F = 4\omega^2\left(ma\omega{u^2} + \frac{\gamma{a}}{ma\omega}uv  + \frac{v^2}{ma\omega} \right)^2.
\end{align}
When the damping coefficient vanishes, $\gamma = 0$, this expression simplifies to:
\begin{align}
\mathcal{I}^{C}_F =   4\omega^2\left(ma\omega{u^2}  + \frac{v^2}{ma\omega} \right)^2,
\end{align}
the Fisher information for the simple harmonic oscillator.

\subsection{The relation between Fisher information and potential energy for the simple harmonic oscillator}\label{SM:Fisher-alternatives}
\setcounter{equation}{0}
For the simple harmonic oscillator, the solutions to the equations of motion are:
\begin{equation}
\begin{aligned}
Q(t) &= B\cos(\omega{t} + \phi) \\
P(t) &= -ma\omega{B}\sin(\omega{t} + \phi).
\end{aligned}
\end{equation}
The time-derivatives are:
\begin{equation}
\begin{aligned}
\dot{Q}(t) = -B\omega\sin(\omega{t} + \phi) \\
\dot{P}(t) = -ma\omega^2{B}\cos(\omega{t} + \phi).
\end{aligned}
\end{equation}
Then, the coordinates of the pure state $\ket{\delta{\yu}_{\dot{\ex}}}$ in the flow direction take the form
\begin{align}\label{eq:unite-less-coordinates-pure-state}
\ket{\delta{\yu}_{\dot{\ex} }}= \begin{pmatrix}
u\\
v\end{pmatrix} = \frac{[\dot{Q},\dot{P}]^\top}{\sqrt{\dot{Q}^2 + \dot{P}^2}} 
= \frac{ [\dot{Q},\dot{P}]^\top}{\norm{\dot{\boldsymbol{X}}}}
\end{align}
where $\norm{\dot{\boldsymbol{X}}} = \sqrt{\dot{Q}^2 + \dot{P}^2}$.
Now, we substitute the expressions of $u$ and $v$ into the Fisher information:
\begin{equation}
\begin{aligned}
\mathcal{I}^C_F(\dot{{\ex}})&= 4\omega^2\left(ma\omega{u}^2 + \frac{v^2}{ma\omega}\right)^2\\
&=4\omega^2 \left(\frac{ma\omega}{1 + \left((ma\omega)^2 -1\right)\cos^2(\omega{t} + \phi)}\right)^2.
\end{aligned}
\end{equation}
The total energy $E$ of the simple harmonic oscillator, the sum of the kinetic energy $K$ and the potential energy $V$, is conserved:
\begin{align}
E = V_\text{max} = q_m{p_m}ma\omega^2{Q^2_{max}}/2 = q_m{p_m}ma\omega^2B^2/2 
\end{align}
Thus, 
\begin{align}
\frac{V}{E} = \frac{q_m{p_m}ma\omega^2B^2\cos^2(\omega{t} + \phi)/2}{q_m{p_m}ma\omega^2B^2/2} = \cos^2(\omega{t} + \phi).
\end{align}
Substituting this result into the last expression of Fisher information, we get:
\begin{align*}
\begin{split}
\mathcal{I}^C_F(\dot{{\ex}}) 
&=  4\omega^2 \left(\frac{ma\omega}{1 + \left(\left(ma\omega\right)^2 -1\right)\frac{V}{E} }\right)^2.
\end{split}
\end{align*}

\subsection{Fisher information bounds for the simple harmonic oscillator}\label{SM:Fisher-bounds}
\setcounter{equation}{0}
One way to find the bounds of $\mathcal{I}^C_F$ is using the purity of the state $\ket{\delta{u}}$, $u^2 + v^2 =1$, with $-1\leq{u}\leq{1}$ and $-1\leq{v}\leq{1}$. The maximum and the minimum of the Fisher information,
\begin{equation}
\begin{aligned}
\mathcal{I}^C_F &=4\omega^2\left(ma\omega + \left(\frac{1}{ma\omega} - ma\omega\right)v^2\right)^2\\
&=4\omega^2\left(\frac{1}{ma\omega} + \left(ma\omega - \frac{1}{ma\omega}\right)u^2\right)^2,
\end{aligned}
\end{equation}
depend on the nondimensionalization parameter $a$:
\begin{align}  
4(ma{\omega^2)^2}\leq{\mathcal{I}^M_{T}}\leq\frac{4}{(ma)^2} \quad\quad a\leq{\frac{1}{m\omega}},\\
\frac{4}{(ma)^2}\leq{\mathcal{I}^M_{T}}\leq4(ma{\omega^2)^2} \quad\quad a\geq{\frac{1}{m\omega}}.
\end{align}  
Another way to find the bounds is by taking the first derivative of $\mathcal{I}^C_F$ with respect to time:
\begin{equation}
\begin{aligned}
\frac{d\mathcal{I}^C_F}{dt} &= 4\frac{d }{dt}\norm{\ket{\delta{\dot{\yu}}}}^2\\
&= 4\left(\bra{\delta{\ddot{\yu}}}\ket{\delta{\dot{\yu}}} + \bra{\delta{\dot{\yu}}}\ket{\delta{\ddot{\yu}}}\right)\\
&= 8 \bra{\delta{\dot{\yu}}}\ket{\delta{\ddot{\yu}}}.
\end{aligned}
\end{equation}
Because $\langle\stability^2\rangle = -\omega^2I$ for the simple harmonic oscillator, where $I$ is the $2\times2$ identity matrix, we get
\begin{align}
\frac{d\mathcal{I}^C_F}{dt} =  -4\langle\stability\rangle\mathcal{I}^C_F.
\end{align}
The maxima and the minima of Fisher information occur when 
\begin{align}
\langle\stability\rangle =  \left(\frac{1}{ma\omega} - ma\omega\right)uv =0.
\end{align}
This condition is satisfied when the potential energy is maximal, which is when there is no momentum $p = P = 0$, and $u=0$ and $v = 1$ in Eq.~\ref{eq:explicit-FI}.
Thus, 
\begin{align}
\mathcal{I}^{C}_F{\left(p=0\right)} = \frac{4}{(ma)^2} = 4\lambda_\text{2}.
\end{align}
The condition is also satisfied when the kinetic energy is maximal, which is when $q =Q=0$ and $u = 1$ and $v = 0$. 
In this case, the potential energy $ V = {q_m{p_m}}ma\omega^2Q^2/2 = 0$. Thus,
\begin{align}
\mathcal{I}^{C}_F{\left(q=0\right)} = 4(ma\omega^2)^2= 4\lambda_\text{1}.
\end{align}
Since the Fisher information has the form
$\mathcal{I}^C_F = 4\left<\stability^\top\stability\right> = 4\Tr(\stability^\top\stability\brho)$, we can use Eq.~\ref{eq:FI-bound00}.
Therefore, 
\begin{align}
4\lambda_\text{min}(\stability^\top\stability)\leq\ \mathcal{I}^{C}_F\leq4\lambda_\text{max}(\stability^\top\stability)
\end{align}
the Fisher information is in the interval $[4\lambda_\text{min},4\lambda_\text{max}]$.

\subsection{Fisher information in terms of the curvature}\label{SM:F-curvature}
\setcounter{equation}{0}
From Eq.~\ref{eq:explicit-FI}, using the coordinates of the pure state in Eq.~\ref{eq:unite-less-coordinates-pure-state}, the equations of motion in Eq.~\ref{eq:unit-less-SHO} and the definition of speed  $\norm{\dot{\boldsymbol X}} = \sqrt{\dot{Q}^2 + \dot{P}^2}$, we get
\begin{align} \nonumber
\label{eq:last}
\mathcal{I}^C_F(\dot{{\ex}})&= 4\omega^2\left(ma\omega{u}^2 + \frac{v^2}{ma\omega}\right)^2\\\nonumber
&= 4\frac{\omega^2}{\norm{\dot{\boldsymbol X}}^4}\left(ma\omega\dot{Q}^2 + \frac{\dot{P}^2}{ma\omega}\right)^2\\\nonumber
&=  4\frac{\omega^6}{\norm{\dot{\boldsymbol X}}^4}\left(\frac{P^2}{ma\omega}+  ma\omega{Q^2}\right)^2\\
&= 16\frac{\omega^6}{\norm{\dot{\boldsymbol X}}^4}
\left(\frac{\mathcal{H}}{{q_m}{p_m}\omega}\right)^2
\end{align}
The last equality uses the Hamiltonian in Eq.~\ref{eq:Hamiltonian-a}.
Because arc length, $s = s\left(Q,P\right)$, is dimensionless here, the derivative $\dot s = \norm{\ket{\dot \ex}} = \norm{\dot {\boldsymbol X}}$ has the units $[T^{-1}]$, which makes the curvature $\kappa$ dimensionless.
Rearranging Eq.~\ref{eq:F-curvature}, we get $\norm{\dot{\boldsymbol X}}^2 = \mathcal{I}^C_F(\dot{{\ex}})/{4\kappa^2}$,
and substituting into Eq.~\ref{eq:last} gives:
\begin{align}
\mathcal{I}^C_F(\dot{{\ex}})  =  \left(16\omega^2\frac{\mathcal{H}}{{q_m}{p_m}}\right)^\frac{2}{3}\kappa^\frac{4}{3}.
\end{align}

\subsection{Relation between state-space curvature and period}
\setcounter{equation}{0}
\label{SM:Period}
From  Eq.~\ref{eq:F-curvature} and the derivative of the arc length $\dot s  = \norm{\dot \ex}$, we get
\begin{align}
\dot s  = \frac{ds}{dt} = \norm{\dot \ex} = \sqrt{\frac{ \mathcal{I}^C_F(\dot{{\ex}})}{4\kappa^2}}
\end{align}
The period is:
\begin{align}\nonumber
T&=\int_{0}^{T} dt = \int_{s_0}^{s}\frac{ds'}{\sqrt{\frac{ \mathcal{I}^{M}_T}{4\kappa^2}}}\\ &=2 \int_{s_0}^{s}\kappa(s')\frac{1}{ \sqrt{\mathcal{I}^{C}_F(\dot{\ex})}} ds'.
\end{align}
For the simple harmonic oscillator when $a = 2$, the Fisher information $\mathcal{I}^{C}_F(\dot{\ex}) = 4\omega^2$, curvature, and radius $\kappa = R^{-1}$ are all constant. The arc length of the circle can be expressed in terms of the angle such that $s = R\theta$ and $ds = Rd{\theta}$. So, the period evaluates to
\begin{align}
  T = 2 \int_{0}^{2\pi}\frac{1}{R}\frac{1}{ \sqrt{4\omega^2}}Rd{\theta} =\frac{2\pi}{\omega},
\end{align}
the well-known result.

\subsection{Expressions of mechanical Fisher information in the flow direction  for the underdamped and critically damped harmonic oscillator}\label{SM:Fisher-under-critically-damped}
\setcounter{equation}{0}
From the solutions $\dot{Q}= \frac{P}{ma}$ and $ \dot{P}=-ma{\omega^2}Q - \frac{\gamma}{m}P$, the Fisher information in the flow direction is:
\begin{align}
\begin{split}
\mathcal{I}^C_F(\dot{{\ex}}) &=   4\omega^2\left(ma\omega{u^2} + \frac{\gamma{a}}{ma\omega}uv  + \frac{v^2}{ma\omega} \right)^2\\
&= 4\frac{\omega^6}{\norm{\dot{\boldsymbol X}}^4}\left(ma\omega{Q^2} + \frac{\gamma{a}}{ma\omega}QP + \frac{P^2}{ma\omega}\right)^2.
\end{split}
\end{align}
For the underdamped harmonic oscillator, substituting the solutions 
\begin{align}
Q(t) &=  C\exp(-\frac{\gamma}{2m}t)\cos(\omega_d{t} + \phi), \\\nonumber
P(t) &= - Ce^{-\frac{\gamma}{2m}t}\left(\frac{\gamma{a}}{2}\cos(\omega_d{t} + \phi) + ma\omega_d\sin(\omega_d{t} + \phi)\right)
\end{align}
into Eq.~\ref{eq:Final-damped-FI} gives the Fisher information
\begin{align}
\mathcal{I}^C_F(\dot{{\ex}}) =  4\omega^2 \left(\frac{ma{\omega}{{{\omega^{2}_d}}}{C^2}e^{-\frac{\gamma}{m}t}}{\norm{\dot{\boldsymbol{X}}}^2}\right)^2.
\end{align}
For the critically damped harmonic oscillator, substituting the solutions
\begin{align}\nonumber
Q(t) &= \left(B + Ct\right)e^{-\frac{\gamma}{2m}t},\\
P(t) &=  \left(maC -\frac{\gamma{a}}{2}B + \frac{\gamma{a}}{2}Ct\right)e^{-\frac{\gamma}{2m}t}.
\end{align}
into Eq.~\ref{eq:Final-damped-FI} gives
\begin{align}
\mathcal{I}^C_F(\dot{{\ex}}) =4\omega^2 \left(\frac{\sqrt{ma{\omega}}{\boldsymbol{C}}e^{-\frac{\gamma}{2m}t}}{\norm{\dot{{{\boldsymbol{X}}}}}}\right)^4.
\end{align}

\section*{Data Availability}

The data that support the findings of this study are available from the corresponding author upon reasonable request.

\section*{References}
\bibliography{references}

\end{document}